\documentclass[12pt]{article}
\usepackage{graphicx}
\newcommand{\bb}{\begin{equation}}
\newcommand{\ee}{\end{equation}}
\newcommand{\ba}{\begin{eqnarray}}
\newcommand{\ea}{\end{eqnarray}}

\begin{document}

\title{{\bf Light Black Holes from Light}}

\author{
Don N. Page
\thanks{Internet address:
profdonpage@gmail.com}
\\
Department of Physics\\
4-183 CCIS\\
University of Alberta\\
Edmonton, Alberta T6G 2E1\\
Canada
}

\date{2025 May 21}

\maketitle
\large
\begin{abstract}
\baselineskip 20 pt

\'{A}lvarez-Dom\'{\i}nguez, Garay, Martin-Mart\'{\i}nez, and Polo-G\'{o}mez \cite{Alvarez-Dominguez:2024pub} have suggested that ``it is not possible to concentrate enough light to precipitate the formation of an event horizon. We argue that the dissipative quantum effects coming from the self-interaction of light (such as vacuum polarization) are enough to prevent any meaningful buildup of energy that could create a black hole in any realistic scenario,'' and ``the dissipation of energy via Schwinger effect alone is enough to prevent the formation of kugelblitze with radii ranging from $10^{-29}$ to $10^8$ m.''  While I agree that it is indeed highly implausible that black holes will form mainly from light in our actual universe, either naturally or by any foreseeable human activity, there are many idealized theoretical processes for forming black holes of any size down to near the Planck length (about $10^{-35}$ m) purely from photons, such as from colliding approximately plane-wave pulses, with only a small fraction of the energy escaping the black hole as scattered light or electron-positron pairs.

\end{abstract}

\normalsize

\baselineskip 22 pt

\newpage

\section{Introduction}

\'{A}lvarez-Dom\'{\i}nguez, Garay, Martin-Mart\'{\i}nez, and Polo-G\'{o}mez \cite{Alvarez-Dominguez:2024pub} (see also \cite{Ball:2024aqz,Loeb:2024kui,Alvarez-Dominguez:2024mjd,Blas:2024ypz,Halilsoy:2024imw}) have noted that if one has a collapsing gas of photons of mass less than about $10^5 M_\odot$ [which in Planck units is about $1.8\times 10^{43}$, close to $1.6\times 10^{43} \approx q/(\pi m^2)$ that is one definition, when $\pi$ is included, of the reciprocal of the critical electric field for the pair production of electrons and positrons of charge magnitude $q = \sqrt{\alpha} \sim \sqrt{1/137}$ and mass $m \approx 4.2\times 10^{-23}$ in Planck units], when it gets sufficiently compacted that it would form a black hole if the photons interacted purely gravitationally, actually the production of electron-positron pairs will lead to the dissipation of the energy before a black hole can form.  Under realistic situations in our universe, such as their assumption of photons traveling largely in random directions at each location, their conclusion seems to be valid.

However, I shall argue that their assumptions and conclusions are not so universally valid as many of their statements suggest.  Instead, there certainly exist idealized situations (initial states with virtually all of the energy in photons) which can form black holes of any mass greater than approximately the minimum mass for a black hole, which is presumably near the Planck mass (about $2.2\times 10^{-8}$ kg or $1.1\times 10^{-38} M_\odot$), with a minimum size near the Planck length (about $1.6\times 10^{-35}$ m), nearly 43 orders of magnitude smaller than the lower limit proposed by \'{A}lvarez-Dom\'{\i}nguez, Garay, Martin-Mart\'{\i}nez, and Polo-G\'{o}mez \cite{Alvarez-Dominguez:2024pub}.

To allow the photon energy to collapse into a black hole, the main situation to be avoided is for many different photons to be traveling in significantly different directions at the same location to produce many electron-positron pairs, until the photon energy gets concentrated enough to form a black hole that engulfs the electron-positron pairs and prevents them from escaping.  For slowly varying classical electromagnetic fields, one generally needs to keep a combination of the Lorentz invariants $\mathbf{E}^2 - \mathbf{B}^2$ and $\mathbf{E}\cdot\mathbf{B}$ from getting near the Schwinger critical value, say $\sqrt{\mathbf{E}^2 - \mathbf{B}^2} = m^2/q$ for $m$ and $q$ the mass and charge of the positron (now for brevity not including a factor of $\pi$), for the rapid production of electron-positron pairs before the energy becomes sufficiently concentrated that it forms a black hole.  These invariants are zero for plane electromagnetic waves, and if one keeps the energy nearly in plane waves (the photons traveling in nearly the same direction) in each region before that region becomes engulfed by a black hole, significant pair production to disperse the energy before it forms a black hole can be avoided.

In the limit that one has a single photon, since it will be in a superposition of plane wave states, each of which will not interact with itself to form electron-positron pairs, the superposition state will also not form pairs.  Of course, such a one-photon quantum state also cannot form a black hole.  However, if one takes a two-photon state with total center-of-momentum (COM) energy $M$ sufficiently larger than the minimum needed to form a black hole, then their collision can form a black hole, with a cross section that will be of the order of $M^2$ in the Planck units that I am using,
\bb
\hbar = c = G = 4\pi\epsilon_0 = 1
\label{Planck}
\ee

In contrast, the Breit-Wheeler cross section \cite{Breit:1934did,Breit:1934zz} for the formation of electron-positron pairs (each particle with the electron mass $m$ and with charge magnitude $q = \sqrt{\alpha}$, using $q$ instead of the more common $e$ for the elementary charge to avoid confusion with the $e$ that I use many times below as the exponential of 1) for $M^2 \sim 1$ is $\approx 16\pi\alpha^2\ln{(M/m)}M^{-2} \sim 0.14\, M^{-2}$, which is less than the cross section to form a black hole for $M > 1$.  Thus a pair of photons with COM total energy $M$ sufficiently larger than the Planck energy will be significantly more probable to form a black hole than an electron-positron pair.  

Furthermore, since when $M \gg 1$, in the COM frame the wavelength of each of the two highly energetic photons, being of the order of $1/M \ll 1$, is small enough so that in principle one can focus the two photons into a region smaller than the size $\sim M$.  Thus the collapse into a black hole can not only be much more probable than pair production, but also can have a probability near unity.

When one considers the case of many photons, each of angular frequency or energy $\omega$ much less than the Planck energy (1 in my units here), the wavelength $2\pi/\omega \gg 1$ puts a limit on how small a region the photons can be focused into before gravitational collapse can take over.  This limit gives an order-of-magnitude limit $M > 1/\omega$ for the mass and hence linear size of the black hole, and therefore a minimum number of photons $N > M/\omega > 1/\omega^2$ of angular frequency $\omega$ that can be focused to form a black hole.

One can now take these $N$ photons, each of angular frequency $\omega \stackrel{>}{\sim} 1/\sqrt{N}$, and put them into two truncated approximately parallel plane wave pulses approaching each other, with characteristic pulse length of the order of $L$ along the direction of propagation (say along the $z$-axis) and radial size $r$ in the transverse directions (parallel to the $x$-$y$ plane, so $\sqrt{x^2+y^2}<r$), obeying the following inequalities:
\bb
\omega^{-1} \leq L < r/2 < M = N\omega > \sqrt{N} > \omega^{-1}.
\label{ineq}
\ee

There is no obstruction to approximately saturating these inequalities and still avoiding dissipating the energy into electron-positron pairs that escape the black hole, so for a number of photons $N$ that is not too large, one can form a black hole of mass $M$ not much larger than $\sqrt{N}$ and hence not much larger than the Planck mass.

\section{Colliding Antiparallel Polarized Plane Wave Pulses}

To give a concrete model for forming a black hole of mass of any mass larger than some lower limit near the Planck mass, I shall put the electromagnetic energy into two colliding pulses, propagating in opposite directions, parallel and antiparallel to the $z$-axis, with the pulses being plane waves parallel to the $x$-$y$ plane for radial distances $\sqrt{x^2+y^2} < r$ from the $z$-axis, but with the energy density dropping off for greater transverse distances so that most of the energy is within the plane-wave region of radius $r$.  To minimize the pair production (though this is not really necessary), I shall take the two pulses to be antiparallel polarized, so that the electric field of each pulse has the same direction throughout that pulse, but has the opposite direction as that of the other pulse, with the invariant $\mathbf{E}^2 - \mathbf{B}^2 < 0$ everywhere that the pulses overlap, and with nowhere a frame in which there is a pure electric field and no magnetic field.

In particular, for $\sqrt{x^2+y^2} < r$, I shall assume that the pulse moving in the positive $z$-direction is a plane wave with electromagnetic field components depending only on the null coordinate $u = t - z$ with the following gaussian profile having maximum electric field magnitude $E_0$ and characteristic length parameter $L$ (so that if one defines a characteristic angular frequency by $\omega^2 = -(1/E_x)(\partial^2/\partial t^2) E_x$  at $t=z$ where the electric field has its maximum value, then $\omega = 1/L$):
\bb
E_x(u) = B_y(u) = E_0 \exp{\left(-\frac{u^2}{2L^2}\right)}
= E_0 \exp{\left(-\frac{(t-z)^2}{2L^2}\right)}.
\label{+z}
\ee
Similarly, I shall assume that also for $\sqrt{x^2+y^2} < r$, the pulse moving the negative $z$-direction is a plane wave with electromagnetic field components depending only on the other null coordinate $v = t + z$ with the corresponding gaussian profile:
\bb
-E_x(v) = B_y(v) = E_0 \exp{\left(-\frac{v^2}{2L^2}\right)}
= E_0 \exp{\left(-\frac{(t+z)^2}{2L^2}\right)}.
\label{-z}
\ee
Therefore, the total electric field of both counter-propagating pulses together is
\bb
E_x = E_0 \exp{\left(-\frac{(t-z)^2}{2L^2}\right)}
-E_0 \exp{\left(-\frac{(t+z)^2}{2L^2}\right)},
\label{E}
\ee
and the total magnetic field of both pulses together is
\bb
B_y = E_0 \exp{\left(-\frac{(t-z)^2}{2L^2}\right)}
+E_0 \exp{\left(-\frac{(t+z)^2}{2L^2}\right)}.
\label{B}
\ee

This electromagnetic field has the invariants $\mathbf{E}\cdot\mathbf{B} = 0$ and
\bb
\mathbf{E}^2 - \mathbf{B}^2 = -4E_0^2 \exp{\left(-\frac{u^2+v^2}{2L^2}\right)}
= -4E_0^2 \exp{\left(-\frac{t^2+z^2}{L^2}\right)}.
\label{inv}
\ee
These mean that at each location in spacetime there is a frame in which only the magnetic field is nonzero, of magnitude
\bb
B = \sqrt{\mathbf{B}^2 - \mathbf{E}^2} = 2E_0 \exp{\left(-\frac{u^2+v^2}{4L^2}\right)}
= 2E_0 \exp{\left(-\frac{t^2+z^2}{2L^2}\right)}.
\label{Binv}
\ee
An observer at rest in that frame at that location is moving along the $z$-axis with rapidity $\psi = tz/L^2$ in the COM frame of the two pulses.

Because at each event there is a frame in which the electric field is zero, the locally constant field approximation (LCFA) gives zero pair production for this field.  However, it is plausible that derivatives of the electromagnetic field may lead to pair production even though it would not if the field were constant at the value it has at any location.  Therefore, I shall show that even for a very conservative upper limit for the pair production, black holes of masses not much larger than the Planck mass can form and engulf almost all possible pair production and prevent the energy from dissipating before gravitational collapse occurs.

The energy density $\rho$ of the two pulses together (even when overlapping) in their joint COM frame is the sum of their separate energy densities,
\bb
\rho = 
\frac{E_0^2}{4\pi}\left(e^{-\frac{u^2}{L^2}} + e^{-\frac{v^2}{L^2}}\right)=
\frac{E_0^2}{4\pi}\left(e^{-\frac{(t-z)^2}{L^2}} + e^{-\frac{(t+z)^2}{L^2}}\right).
\label{rho}
\ee

The integral over the entire $z$-coordinate (at any fixed $x$ and $y$ for this plane-wave pulse) of this energy density gives a mass $M$ per cross-sectional area $A$ (in the $x$-$y$ plane) of
\bb
\frac{M}{A} = \frac{E_0^2 L}{2\sqrt{\pi}}.
\label{M/A}
\ee
If at $t=0$ one restricts the integral over $z$ to $-L < z < L$, the integral gives an amount less by a factor of the error function of 1, which is approximately 0.8427.  One might note that if one restricts the integral over $z$ to run at least from $-0.55L$ to $+0.55L$, one gets $M/A > E_0^2 L/(2\pi)$.

A conservative sufficient condition for a black hole to form at $t=0$ (when the two pulses completely overlap, giving a pure magnetic field in the COM frame), is that the energy $M$ in the COM frame inside a sphere of radius $R = \sqrt{x^2+y^2+z^2}$ gives a Schwarzschild radius $2M > R$, or $2M/R > 1$.  The integral of the energy density $\rho$ over all $x^2+y^2+z^2 < R^2$ at $t=0$, divided by $R/2$, gives, with 
\bb
F \equiv \frac{R}{L},
\label{F}
\ee
\bb
\frac{2M}{R} = E_0^2 L^2 K(F) \equiv \epsilon K(F),
\label{2M/R}
\ee
where
\bb
\epsilon \equiv E_0^2 L^2,
\label{epsilon}
\ee
and
\bb
K(F) \equiv \left(F - \frac{1}{2F}\right)\sqrt{\pi}\,\mathrm{erf}(F) + e^{-F^2}
= \frac{4}{3}F^2 - \frac{4}{15}F^4 + \frac{2}{35}F^6 - \frac{2}{189}F^8 + \frac{1}{594}F^{10} + O(F^{12}),
\label{K}
\ee
with error function $\mathrm{erf}(F) \equiv (2/\sqrt{\pi})\int_0^F e^{-w}dw$, so that for $R \ll L$ or $F \ll 1$, one has $2M/R \approx (2/R)(4\pi/3)R^3\rho_0$, where $\rho_0 = E_0^2/(2\pi)$ is the energy density at $t=z=0$ where the total electric field is zero and the total magnetic field is $2E_0$.

At the other extreme, for $R \gg L$, or $F \gg 1$, which is the relevant extreme for which negligible energy escapes as electron-positron pairs rather than forming a black hole, neglecting $e^{-F^2}$, and hence also $1-\mathrm{erf}(F)$, gives $K(F) \approx \sqrt{\pi}\,[F-1/(2F)]$ so that
\bb
\frac{2M}{R} \approx \frac{E_0^2}{R}\int_{-\infty}^{+\infty}(R^2-z^2)e^{-\frac{z^2}{L^2}} dz = \sqrt{\pi}E_0^2\frac{L}{R}(R^2-\frac{1}{2}L^2)
= \sqrt{\pi} E_0^2 L R \left(1 - \frac{1}{2 F^2}\right).
\label{2M/RlargeF}
\ee

The minimum mass, say $M_0$, for a black hole that can form from the collision of the two plane wave pulses with parameters $E_0$ and $L$ would be approximately $M_0 = R/2 = (L/2) F$, where $F \equiv R/L$ takes the value that makes $2M/R = E_0^2 L^2 K(F) = \epsilon K(F) = 1$, where $K(F) = 1/\epsilon = 1/(E_0L)^2$.

An approximate inversion of the formula $K(F) = 1/\epsilon$ for all positive values of $\epsilon = E_0^2 L^2$ that matches $K(F) \approx (4/3)F^2 - (4/15)F^4$ for $F \ll 1$ and $K(F) \approx \sqrt{\pi}\,[F-1/(2F)]$ for $F \gg 1$ is
\bb
F(\epsilon) \approx \hat{f}(\epsilon) = \frac{1}{\sqrt{\pi}\epsilon}
\sqrt{\frac{1+\beta\epsilon+\gamma\epsilon^2+\delta\epsilon^3}
{1+\beta\epsilon+(\gamma-\pi)\epsilon^2}},
\label{fhat}
\ee
where (leaving the letter $\alpha$ for the fine-structure constant that is the square of the elementary charge $q=e$, $\alpha = q^2 \approx 1/137$)
\ba
\beta &=& \frac{67\,200\pi-8\,964\pi^2}{81\pi^2-20\,160\pi+89\,600} \approx 4.531\,469\,033\,83, \nonumber \\
\gamma &=& \frac{81\pi^3 + 40\,320\pi^2}{81\pi^2-20\,160\pi+89\,600} \approx 14.796\,046\,284\,8, \nonumber \\
\delta &=& \frac{45\,360\pi^3 - 67\,200\pi^2}{81\pi^2-20\,160\pi+89\,600} \approx 27.460\,159\,432\,0, \nonumber \\
\beta-\pi &=& \frac{60\,480\pi^2 - 89\,600\pi}{81\pi^2-20\,160\pi+89\,600} \approx 11.654\,453\,631\,2.
\label{coefficients}
\ea

For $\epsilon \equiv (E_0 L)^2 < 0.2$, which is the regime relevant for negligible dissipation by electron-positron pairs, a simpler approximation for $F(\epsilon)$ with less than 1\% relative error for such small values of $\epsilon$ is
\bb
f(\epsilon) = \frac{1}{2}\left(\frac{1}{\sqrt{\pi}\epsilon} + \sqrt{\frac{1}{\pi\epsilon^2}+2}\right).
\label{f}
\ee
In the case in which $\epsilon \equiv E_0^2 L^2 \ll 1$, one gets the even simpler approximation
\bb
F \equiv \frac{R}{L} \approx \frac{1}{\sqrt{\pi}\epsilon} = \frac{1}{\sqrt{\pi}E_0^2 L^2}.
\label{Fapprox}
\ee

Then the minimum mass black hole that can be formed with the two counter-propagating pulses with parameters $E_0$ and $L$ is approximately
\bb
M_0 \sim \frac{1}{2} R = \frac{1}{2} L F \sim \frac{1}{2} L \hat{f}(E_0^2 L^2) \sim \frac{1}{2} L f(E_0^2 L^2) \sim \frac{L}{2\sqrt{\pi}\epsilon} = \frac{1}{2\sqrt{\pi}E_0^2L},
\label{M0}
\ee
with the approximate function $\hat{f}(\epsilon)$ given by Eq.\ (\ref{fhat}) for all $\epsilon$, or $f(\epsilon)$ given by Eq.\ (\ref{f}) for $\epsilon \stackrel{<}{\sim} 0.2$, with argument $\epsilon \equiv E_0^2 L^2$ in terms of the parameters of the pulses, and with the last two expressions on the right hand side approximately valid for $\epsilon \ll 1$.  A black hole near this minimum mass for each set of the parameters $E_0$ and $L$ will only occur if the transverse width of the pulse is not much greater than $R = L F(E_0^2 L^2)$.  Otherwise, if the pulse has field values roughly $E_0$ extending transversely to $R >> L f(E_0^2 L^2)$, the mass of the black hole will be of the order of $M \sim E_0^2 L R^2 \sim (E_0^4 L^2 R^2) M_0 = (\epsilon F)^2 M_0$.  However, the point is that by making the transverse width not much greater than necessary, one can form a black hole of mass $M \sim L F(E_0^2 L^2)$, which for $\epsilon = E_0^2 L^2 \ll 1$ is $M \sim 1/(E_0^2 L) \sim L/\epsilon$.

One can also show, independently of the approximation $\hat{f}(\epsilon)$ or $f(\epsilon)$ or $F \sim 1/(\sqrt{\pi}\epsilon)$, that if the criterion for a black hole to form is that the energy density $\rho$ given by Eq.\ (\ref{rho}) integrated over the ball of radius $R$ equals $R/2$, then
\bb
M > M_0 = (1/2) L F(E_0^2L^2) > \frac{1}{2\sqrt{\pi}E_0^2L} = \frac{L}{2\sqrt{\pi}\epsilon}.
\label{Mmin}
\ee
However, this approximation essentially uses the hoop conjecture \cite{Klauder:1972je} in flat spacetime and so ignores the spacetime curvature and gravitational focusing of the counter-propagating electromagnetic pulses that might allow black holes to form of lower mass than the flat spacetime hoop conjecture would give.  It also ignores the energy escaping from outside the possible formation of a black hole, which we turn to next.

\section{Conservative Upper Bound on Pair Production}

Now I wish to find conservative upper bounds to the number of electron-positron pairs produced and energy dissipated that can escape the putative black hole, possibly preventing its formation.

For a constant electromagnetic field with the Lorentz invariants $\mathbf{E}^2 - \mathbf{B}^2$ and $\mathbf{E}\cdot\mathbf{B}$, one can define the subsidiary nonnegative invariants
\ba
E &=& \sqrt{\frac{1}{2}\sqrt{(\mathbf{E}^2 - \mathbf{B}^2)^2 + 4(\mathbf{E}\cdot\mathbf{B})^2} + \frac{1}{2}(\mathbf{E}^2 - \mathbf{B}^2)}, \nonumber \\
B &=& \sqrt{\frac{1}{2}\sqrt{(\mathbf{E}^2 - \mathbf{B}^2)^2 + 4(\mathbf{E}\cdot\mathbf{B})^2} - \frac{1}{2}(\mathbf{E}^2 - \mathbf{B}^2)},
\label{EB}
\ea
which are the magnitudes of the electric and magnetic fields in any frame in which $\mathbf{E}$ and $\mathbf{B}$ are parallel or zero.  In such a constant electromagnetic field, the pair-production rate per spatial volume per time of electrons and positrons of charge magnitude $q = e$ and mass $m$ is the first term in the infinite series for twice the imaginary part of the one-loop effective action per 4-volume \cite{Sauter:1931zz,Heisenberg:1936nmg,Weisskopf:1936hya,Schwinger:1951nm,Nikishov:1969tt,Nikishov:1970br},
\bb
\mathcal{N} = \frac{(qE)(qB)}{(2\pi)^2}\coth{\left(\frac{\pi B}{E}\right)}\exp{\left(-\frac{\pi m^2}{qE}\right)}.
\label{Npair}
\ee
In the limit that the invariant scalar $B$ is taken to 0, this becomes
\bb
\mathcal{N} = \frac{q^2E^2}{4\pi^3}\exp{\left(-\frac{\pi m^2}{qE}\right)}.
\label{NpairB0}
\ee

However, for the colliding antiparallel polarized plane wave pulses, the electric and magnetic fields given by Eqs.\ (\ref{E}) and (\ref{B}) give the Lorentz invariant scalar $E=0$ everywhere, so the locally constant field approximation (LCFA) gives $\mathcal{N}=0$ everywhere, predicting no pair production.

Nevertheless, it seems implausible that there would actually be no pair production even though the LCFA gives $\mathcal{N}=0$.  Therefore, I shall take as a conservative upper limit what one would get from Eq.\ (\ref{NpairB0}) if one took the maximum value of 1 for the exponential and in the prefactor replaced the invariant $E$ with the invariant $B$ from Eq.\ (\ref{Binv}), giving
an upper limit for the pair-production rate per 4-volume of
\bb
\mathcal{N_{\mathrm{max}}} = \frac{q^2B^2}{4\pi^3} = \frac{q^2E_0^2}{\pi^3}e^{-\frac{t^2+z^2}{L^2}}.
\label{Nmax}
\ee

Assuming flat spacetime with no black hole formation, this conservative upper limit for the pair production rate per 4-volume would, when integrated over the entire infinite range of both $t$ and $z$, give a maximum number of pairs, $N_{\mathrm{max}}$, per transverse area $A$, of
\bb
\frac{N_{\mathrm{max}}}{A} = \frac{q^2 E_0^2 L^2}{\pi^2} = \frac{\alpha\epsilon}{\pi^2},
\label{Nmax/A}
\ee
where $\alpha = q^2 \approx 1/137$ is the fine structure constant and where Eq.\ (\ref{epsilon}) gives $\epsilon \equiv E_0^2L^2$ in terms of the maximum electric field magnitude $E_0$ of each plane wave gaussian pulse with length parameter $L$ along the propagation direction $\pm z$.

Taking the photon frequency to be $\omega = 1/L$ and using Eq.\ (\ref{M/A}) gives the number of photons per cross sectional area as
\bb
\frac{N_\gamma}{A} = \frac{ML}{A} = \frac{E_0^2 L^2}{2\sqrt{\pi}} = \frac{\epsilon}{2\sqrt{\pi}},
\label{Ngamma/A}
\ee
so the conservative upper limit of the number of pairs produced, divided by the photon number, is
\bb
\frac{N_\mathrm{max}}{N_\gamma} = \frac{2\alpha}{\pi^{3/2}} \approx 0.002621.
\label{Nmax/Ngamma}
\ee

Unless each charged particle can gain more energy than the typical photon energy $\omega = 1/L$, this may already suggest that the pair production has only a small effect on the evolution of the energy in the electromagnetic pulses.  However, it is not completely clear that this is the case, since for a constant electromagnetic field, the individual photon frequency and energy is zero, so the pairs gain an infinite multiple of the individual photon energy.  Of course, since the LCFA gives zero pair production for the two colliding antiparallel polarized plane wave pulses, one might suppose in this case in which only the inhomogeneities in the electromagnetic field can lead to pair production, it might be plausible that the energy of each electron and positron is limited by the effective frequency $1/L$ of the electromagnetic field.

Nevertheless, I shall now show that there are theoretically possible ranges for the $E_0$ and $L$ parameters so that the pair production outside the black hole region is so greatly exponentially suppressed by the $\exp{-(t^2+z^2)/L^2}$ factor outside the region where the black hole is expected to form that it cannot prevent a black hole from indeed forming.

Consider the formation of a black hole of Schwarzschild radius $R$ such that $e^{-F^2} = e^{-R^2/L^2}$ is very small, so that $1/F = L/R$ is also small but need not be extraordinarily small.  At first ignoring the pair production, this black hole will form from the electromagnetic energy inside a cylinder of radius $x^2+y^2 = r^2 \sim R^2$, energy $\pi r^2 M/A = (\sqrt{\pi}/2)r^2 E_0^2 L$, which equals the black hole mass $M = R/2$ for 
\bb
r \sim R \sim \frac{1}{E_0^2 L} \equiv \frac{L}{\epsilon},
\label{R}
\ee 
dropping $O(1)$ constants such as $\pi$.  Within $x^2+y^2 \stackrel{<}{\sim} r^2$, the total number of pairs produced outside the black hole region, $t^2 + z^2 \stackrel{>}{\sim} R^2 \sim r^2$ will be conservatively bounded above by the transverse area $\pi r^2$ multiplying the integral of
$\mathcal{N_{\mathrm{max}}} = \frac{q^2E_0^2}{\pi^3}e^{-\frac{t^2+z^2}{L^2}}$ over this outside region, which gives
\bb
N_\mathrm{outside} \stackrel{<}{\sim} q^2 E_0^2 L^2 R^2 e^{-R^2/L^2} = \alpha\epsilon R^2 e^{-R^2/L^2}.
\label{Noutside}
\ee

Now a conservative upper bound on the energy per pair is $q E_0 R$, which gives a conservative upper bound on the energy of the pairs as 
\bb
\mathcal{E} \stackrel{<}{\sim} q E_0 R N_\mathrm{outside} \stackrel{<}{\sim} q^3 E_0^3 L^2 R^3 e^{-R^2/L^2}.
\label{Epairs}
\ee
When using $M \sim R \sim 1/(E_0^2 L) \equiv L/\epsilon$ from Eq.\ (\ref{R}), this leads to a very conservative upper limit
on the ratio of the total energy of the pairs to the energy of the black hole as
\bb
\frac{\mathcal{E}}{M} \stackrel{<}{\sim} q^3 E_0^3 L^2 R^2 e^{-R^2/L^2} \sim \frac{q^3}{E_0}e^{-R^2/L^2} 
= \frac{q^3}{E_0}e^{-F^2} \sim \frac{q^3 M}{\sqrt{F}}e^{-F^2} \stackrel{<}{\sim} \frac{q^3}{E_0} e^{-1/(\pi\epsilon^2)}.
\label{E/M}
\ee

Therefore, out of the energy that would go into a black hole if there were no pair production, a very conservative upper limit on the fraction of the energy that would go into electron-positron pairs rather than into a black hole is itself very small if (though not only if)
\bb
F \equiv \frac{R}{L} \stackrel{>}{\sim} \frac{1}{\sqrt{\pi}\epsilon} \equiv \frac{1}{\sqrt{\pi}E_0^2 L^2}  \gg \sqrt{\ln{(q^3 M + 2)}},
\label{Flimit}
\ee
where (all in Planck units) $E_0$ is the maximum electric (or magnetic) field of each electromagnetic plane wave gaussian pulse, $L$ is the rms spread of the pulse in the $z$-direction, $R \gg L$ is an approximate radius of a sphere containing the electromagnetic energy that undergoes gravitational collapse to form the black hole, $q = \sqrt{\alpha} \approx 0.0854$ is the positron charge, and $M \approx R/2 \approx 1/(2\sqrt{\pi}E_0^2 L) = L/(2\sqrt{\pi}\epsilon)$ is the mass of the black hole that forms.  (The 2 near the end of the last expression on the right hand side of Eq.\ (\ref{Flimit}) is just a rather arbitrary simple $O(1)$ number chosen to be sufficiently larger than 1 to keep the logarithm bounded below by a number of the order of unity even when $q^3M$ might be small, so that the inequality implies that $F \gg 1$.)

In qualitative terms, only a negligible fraction of the relevant energy gets dissipated into electron-positron pairs if energy $M$ gets within a sphere of radius $R \approx 2M$ before there is any significant overlap of the two electromagnetic pulses, each of length along the $z$-direction of propagation of the order of $L \ll R = FL$.  

As an example, consider a sphere centered on the spatial origin with radius $R = F L$ and with $F \equiv R/L \gg 1$, and consider a time $t = -R/\sqrt{2}$ so that at that time the central plane of each pulse is at $z = \pm R/\sqrt{2}$.  The sphere of radius $R$ encloses disks on the pulse central planes with $x^2+y^2 < r^2 = R^2 - z^2 = R^2/2$.  Since $L^2 \ll R^2$, the total energy within the sphere is, using Eq.\ (\ref{M/A}),
\bb
M \approx \frac{M}{A}\pi r^2 \approx \frac{E_0^2 L}{2\sqrt{\pi}}\pi R^2/2 = \frac{\sqrt{\pi}}{4}E_0^2 L^3 F^2 
= \frac{\sqrt{\pi}}{4} L \epsilon F^2.
\label{Minside}
\ee
For this energy already to be within a black hole, we need $R = FL \stackrel{<}{\sim} 2M$, the Schwarzschild radius of the energy within the sphere of radius $R$, so
\bb
F \equiv \frac{R}{L} \stackrel{>}{\sim} \frac{2}{\sqrt{\pi}\epsilon} = \frac{2}{\sqrt{\pi}E_0^2 L^2}.
\label{F>}
\ee

This then implies that $\epsilon \equiv E_0^2 L^2 \stackrel{>}{\sim} 2/(\sqrt{\pi} F)$, which when inserted back into Eq.\ (\ref{Minside}) gives
\bb
M \stackrel{>}{\sim} \frac{1}{2} F L \stackrel{>}{\sim} \frac{1}{\sqrt{\pi}E_0^2L} = \frac{L}{\sqrt{\pi}\epsilon}.
\label{M>FL/2}
\ee

The approximate lower bound on $F$ in Eq.\ (\ref{F>}) is a factor of 2 larger than the approximation of Eq.\ (\ref{Fapprox}) because that approximation was for the time $t=0$ when the two pulses completely overlap (which gives the maximum mass within the sphere of radius $R$), whereas the present calculation is for $t = -R/\sqrt{2}$ when the central planes of the two pulses are separated by a distance $\sqrt{2}R$, so that the radius $r$ of the disks of the central planes of the two pulses is $R/\sqrt{2}$, and hence have half the area and half the enclosed energy as the disks of radius $r=R$ enclosed by the sphere of radius $R$ at $t=0$, when the central plane of each pulse coincides and is the $x$-$y$ plane, $z=0$.  Of course, in both cases I am taking the approximation of flat spacetime, which will not be valid for the region inside and near the black hole, but this approximation should be sufficient for showing that with $F$ sufficiently large, the pair production can be negligible until the two pulses get inside a black hole, so that their energy cannot escape the black hole.

At time $t = -R/\sqrt{2} = -LF/\sqrt{2}$, the invariant magnetic field magnitude, as given by Eq.\ (\ref{Binv}), is
\bb
B = \sqrt{\mathbf{B}^2 - \mathbf{E}^2} = 2E_0e^{-\frac{1}{4}F^2 - \frac{z^2}{2L^2}} \leq 2E_0 e^{-\frac{1}{4}F^2}
\stackrel{<}{\sim} 2E_0 e^{-\frac{1}{\pi\epsilon^2}} = 2E_0 \exp{\left(-\frac{1}{\pi E_0^4 L^4}\right)}.
\label{Bt}
\ee

In the Planck units I am using, the unit magnetic field in the SI unit of a tesla (T) is the square root of the fine structure constant multiplied by the Planck energy in electron volts and divided by the Planck length in metres and also divided by the speed of light in metres per second, which works out to be $2.152\,428(47)\times 10^{53}$ T.  Analogously, the Planck unit electric field in the SI unit of volts per metre (V/m) is the speed of light in metres per second multiplying the Planck unit magnetic field in tesla, or $6.452\,82(14)\times 10^{61}$ V/m.  

In quantum electrodynamics, the critical electric field in SI units (volts per metre) can be considered to be $E_c = m^2c^3/(\hbar q) = 1.323\,285\,477\,72(82)\times 10^{18}$ V/m, and the critical magnetic field in SI units (tesla) can be considered to be $B_c = m^2c^2/(\hbar q) = 4.414\,052\,306\,5(27)\times 10^9$ T $\approx 4.414$ gigatesla.  In Planck units, both of these can be considered to be $E_c = B_c = m^2/q = m^2/\sqrt{\alpha}$, where $m = 4.185\,463(46)\times 10^{-23}$ is the electron or positron mass in Planck units, and where $q = \sqrt{\alpha} = 0.085\,424\,543\,103(7)$ is the magnitude of the electron or positron charge in Planck units.  This gives a critical electric or magnetic field of $2.050\,710(45)\times 10^{-44}$ in Planck units.  However, some take the critical fields to be $\pi$ times these, or $E_c' = \pi E_c = \pi m^2c^3/(\hbar q) = 4.157\,223\,935\,4(26)\times 10^{18}$ V/m and $B_c' = \pi B_c = \pi m^2c^2/(\hbar q) = 1.386\,700\,640\,55(86)\times 10^{10}$ T $\approx 13.867$ gigatesla.  In both cases in Planck units one gets $E_c' = B_c' = 6.442\,50(14)\times 10^{-44}$.  A motivation for using the larger definition of the critical values is that then the exponential in the pair-production rate becomes $\exp{(-E_c'/E)}$.

Now if in Eq.\ (\ref{Bt}) we choose as an example $F = 24$, then $\epsilon \equiv E_0^2 L^2 \approx 2/(\sqrt{\pi}F) \approx 0.0470$ or $E_0 L = \sqrt{\epsilon} \sim 0.217$, which is fairly near the Planck value, but if one wants $L$ to be at least as large as the Planck length, so $E_0 \stackrel{<}{\sim} 0.217$, then the maximum value of the invariant magnetic field $B$ for $t \leq -R/\sqrt{2}$ is
\bb
B \approx 2E_0 e^{-\frac{1}{4}F^2} \approx 5.789\times 10^{-63}E_0 \approx (1.25\, \mathrm{nT})E_0 \stackrel{<}{\sim} 0.27\, \mathrm{nanotesla}.
\label{Bexample}
\ee
This is a very tiny magnetic field, so that even an electric field of this same strength would be far too small to produce any electron-positron pairs.  Therefore, for this example of $F = R/L = 24$, before any electron-positron pairs are produced, one can produce a black hole of mass $M \stackrel{>}{\sim} (F/2)L = 12L$ that can be within an order of magnitude or so of $L$.  There is no obvious reason why $L$ cannot be as small as the order of magnitude of the Planck length.  (Indeed, I do not see any reason why one could not in principle have $L$ arbitrarily small, as least if there is no limit to the rapidity of a Lorentz transformation, though if $ML$ is small, the expected number of photons forming the black hole, which is approximately $ML$, would be small, so that a classical approximation would not be valid as I would expect it to be for $ML \gg 1$ in the Planck units I am using.)

As a result, an initial state with two nearly plane-wave electromagnetic pulses can form a black hole of mass not much greater than the Planck mass ($2.176\,434(24)\times 10^{-8}$ kg) and size not much larger than the Planck length ($1.616\,255(18)\times 10^{-35}$ m) without significant electron-positron pair production before the black hole forms to dissipate the energy.  (Significant pair production may occur where the pulses strongly overlap inside the black hole, but these pairs cannot escape from the hole.)

Of course, to avoid making a black hole much larger than the minimum size possible, $M_0 \sim 1/(2\sqrt{\pi}E_0^2 L) = L/(2\sqrt{\pi}\epsilon)$ for the pulse parameters $E_0$ (the maximum electric and magnetic field of each pulse) and $L$ (the effective length of each pulse in its direction of propagation), one needs to have a transverse cutoff of the pulses near $R \sim 2M_0 \sim 1/(\sqrt{\pi}E_0^2 L) = L/(\sqrt{\pi}\epsilon)$ so that the pulses much wider than this do not produce black holes of mass $M$ much larger than $M_0$.  The details of such a cutoff is beyond the scope of this paper, but for $F \equiv R/L \stackrel{>}{\sim} 2/(\sqrt{\pi}\epsilon) \equiv 2/(\sqrt{\pi}E_0^2 L^2) \gg 1$, there should be no problem in making this cutoff in a way that keeps the mass close to the minimum that can itself be made not not far from the Planck mass.

Thus indeed one can produce light black holes from light in such a theoretically possible scenario, even though such a scenario is very unlikely to occur in our present universe.

\section{Conclusions}

Although I agree with the strong suggestion of \'{A}lvarez-Dom\'{\i}nguez, Garay, Martin-Mart\'{\i}nez, and Polo-G\'{o}mez \cite{Alvarez-Dominguez:2024pub} that ``the formation of black holes solely from electromagnetic radiation is impossible, either by concentrating light in a hypothetical laboratory setting or in naturally occurring astrophysical phenomena,'' if instead one allows purely theoretical situations within the known laws of physics, it is in principle possible to concentrate enough light to precipitate the formation of an event horizon.  The trick, probably never actually occurring in our universe either naturally or by human intervention, would be to concentrate the radiation into two plane pulses (each of which separately do not produce pairs) that do not overlap enough to give significant pair production before they have concentrated enough energy close enough to each other to form a black hole that would engulf any pair production that occurs when the two pulses do eventually get close enough together to produce many pairs inside the black hole.

\section{Acknowledgments}

I have benefited from an email exchange with Eduardo Martin-Mart\'{\i}nez, even though I have not had time to discuss all the interesting issues he raised.  This research was supported by a grant from the Natural Sciences and Enginneering Research Council of Canada.

\end{document}